\documentclass{aa}  
\usepackage{graphicx}
\usepackage{txfonts}
\usepackage{wasysym}
\usepackage{color}
\begin{document}
%

\title{The spectroscopic orbit and the geometry of R~Aqr\\~\\}

   \author{M. Gromadzki
          \inst{1}
          \and
          J. Miko{\l}ajewska\inst{1}          }

   \offprints{M. Gromadzki}

   \institute{Nicolaus Copernicus Astronomical Center,
              Bartycka 18, 00-716 Warszawa, Poland\\
              \email{marg,mikolaj@camk.edu.pl}
             }

   \date{Received October ??, ????; accepted ???? ??, ????}

 
  \abstract
   {R Aqr is one of the closest symbiotic binaries and the only D-type system with radial velocity data suitable for orbital parameter estimation.}
   {The aims of our study are to derive a reliable spectroscopic orbit of the Mira component, and to establish connections between the orbital motion and other phenomena exhibited by R~Aqr.}
   {We reanalyze and revise the velocity data compiled by McIntosh \& Rustan
complemented by additional velocities.}
   {We find an eccentric orbit ($e=0.25$) with a period 43.6 yr. This solution is in agreement with a
   resolved VLA observation of this system. We demonstrate that the last increase in extinction towards the Mira variable in 1974--1981 occurred during its superior, spectroscopic conjunction, and can be due to obscuration by a neutral material in the accreting stream.  We also show that jet ejection is not connected with the orbital position.
}
   {}


   \maketitle
%

\section{Introduction}

R~Aqr is a symbiotic binary surrounded by an hourglass nebula (Solf \& Ulrich \cite{solf}), and 
both its cool and hot components are unique.
The cool component is a Mira variable with a pulsation period of 387$^{\rm d}$
with SiO and H$_{2}$O maser emission, which is rare, with only three among 48 symbiotics with Miras exhibiting this emission (Whitelock \cite{whitelock-review}).
The hot component shows sporadic, jet ejections ({\rm e.g.} Kellogg et al. \cite{kellogg} and Nichols et al. \cite{nichols}), and
Nichols et al. (\cite{nichols}) also detected a 1734 s periodic oscillation in {\it Chandra} X-ray observations, 
which suggested that the hot companion is a magnetic white dwarf.
Willson et al. (\cite{willson}) interpreted a brightness reduction in pulsation maxima as the eclipsing of the Mira by a companion surrounded by an extended gas cloud.
This interpretation was supported by two facts. First, during the last hypothetical eclipse (1974--1980), a minimum was observed in the near-IR light
curve similar to that corresponding to dust obscuration in other symbiotic Miras ({\rm e.g.}, Whitelock et al. \cite{whitelock1983}). Secondly, a spectrum obtained during the first such event (1886--1894) exhibits only faint emission lines of hydrogen  and no trace of M type spectral features (Townley et al. \cite{townley}). 
In the historical, visual, light curve, three phases of reduced maxima are present spaced at 44 yr intervals and
this value was adopted by Willson et al. (\cite{willson}) as the orbital period. Additionally, the $O-C$ diagram showed variation with a period of $\sim$22 yr.
Wallerstein (\cite{wallerstein}) used a number of spectral absorption lines associated with the Mira in an attempt
to determine the orbital elements of R~Aqr. His results were consistent with the 44 yr period, but uncertain. 
Hinkle et al. (\cite{hinkle}) combined these archival data with  radial velocities (RV) determined from
near-IR CO and Ti I lines, and obtained a 44 yr period and a highly eccentric, $e=0.6$, orbital solution.

McIntosh \& Rustan (\cite{mcintosh}) collected radial velocity data of SiO maser emission of R Aqr (Lepine et al.
    \cite{lepine}; Cohen \& Ghigo \cite{cohen}; Spencer et al. \cite{spencer}; Lane \cite{lane}; Hall et al. \cite{hall}; Cho et al. \cite{cho}; 
    Boboltz \cite{boboltz}; Hollis et al. \cite{hollis}; Alcolea et al. \cite{alcolea}; Pardo et al. \cite{pardo};  McIntosh \& Rustan \cite{mcintosh}) complemented by previous radial velocities derived from optical absorption lines (Merrill \cite{merrill2}; Jacobsen \& Wallerstein \cite{jacobsen}; Wallerstein \cite{wallerstein}) and near-IR CO and Ti I lines (Hinkle et al. \cite{hinkle}). All of these features are associated with the cool component or material in its close neighborhood and reflect its motion around the mass center. 
In astrophysical masers, emission originating in many maser spots moves
around the Mira variable. Spectra of masers exhibit a series of emission lines with 
various velocities associated with different maser spots. 
 McIntosh
(\cite{mcintosh06}) studied changes in the velocity centroid (VC) of SiO maser emission from 76 single evolved giants, and showed that the mean difference in the VCs measured at different epochs was only $0.065 \pm 2.00\, \rm km\,s^{-1}$, and that the VC of the emission distribution 
can be used to derive the velocity of the star. Based on this result, McIntosh \& Rustan (\cite{mcintosh}) adopted the VC in their 
work for all kinds of velocity data analysis. 

McIntosh \& Rustan 
(\cite{mcintosh}) also estimated orbital elements using a program developed by Gudehus (\cite{gudehus}). 
They inferred an eccentric orbit, $e=0.52$, with a period 34.6 yr.  Their orbital period was significantly 
shorter than any previous estimates as well as the period derived in our present study. Their other were of reasonable accuracy
if we interpret their system mass, equal to 0.043  $\rm M_{\odot}$,
as the mass function, and their semi-major axis of the system, of 3.7 AU,
as the semi-major axis of the orbit of the Mira component.
We also note that not all points listed in their Table 1 were plotted on their  figures showing VC versus orbital phase, namely, they  did not measurement from Alcolea et al. (\cite{alcolea}). 
In the text, the authors wrote that points from Hollis et al. (\cite{hollis}) are listed in their Table 1, although they were not displayed there. These data points were plotted 
on the figures showing VC versus Julian days, but
they were missing on the figure of VC versus orbital phase.  
The authors also did not compare the orbit solution with the resolved image of the binary (Hollis et al. \cite{hollis2}), nor include radial velocity measurements before 1946 (Merrill \cite{merrill1}, \cite{merrill2}).
These points were also ignored in all previous papers about the orbit estimation from radial
velocities (Wallerstein \cite{wallerstein}, Hinkle et al. \cite{hinkle}).
The authors also neglected radial velocity  measurements of the SiO maser by 
Zuckerman et al. (\cite{zuckerman}), Hollis et al. (\cite{hollis3}), Hall et al. (\cite{hall1}), Allen et al. (\cite{allen}), Heske (\cite{heske}), Jewell et al. (\cite{jewell}), Patel et al. (\cite{patel}), Gray et al. (\cite{gray}), Schwarz et al. (\cite{schwarz}), 
Hollis et al. (\cite{hollis}, \cite{hollis4}), Imai et al. (\cite{imai}), Cotton et al. (\cite{cotton1}, \cite{cotton2}), and Kang et al. (\cite{kang}).
We also note that the observations studied by Pardo et al. (\cite{pardo}) included the observations analyzed by
Alcolea et al. (\cite{alcolea}), which included the data used by Martinez et al. (\cite{martinez}). These three data sets are thus not entirely different, and the later studies were based simply on data collected over longer time.

This paper contains an analysis of the revised radial velocity data
listed in Table 1 of McIntosh \& Rustan (\cite{mcintosh}) and additional
velocity data (listed here in Table \ref{table:1}),
as well as discussion of the connections between the orbital motion
and other phenomena shown by R Aqr.


\section{Orbit determination}

  \begin{figure*}
   \centering
   \includegraphics[angle=-90,width=\textwidth]{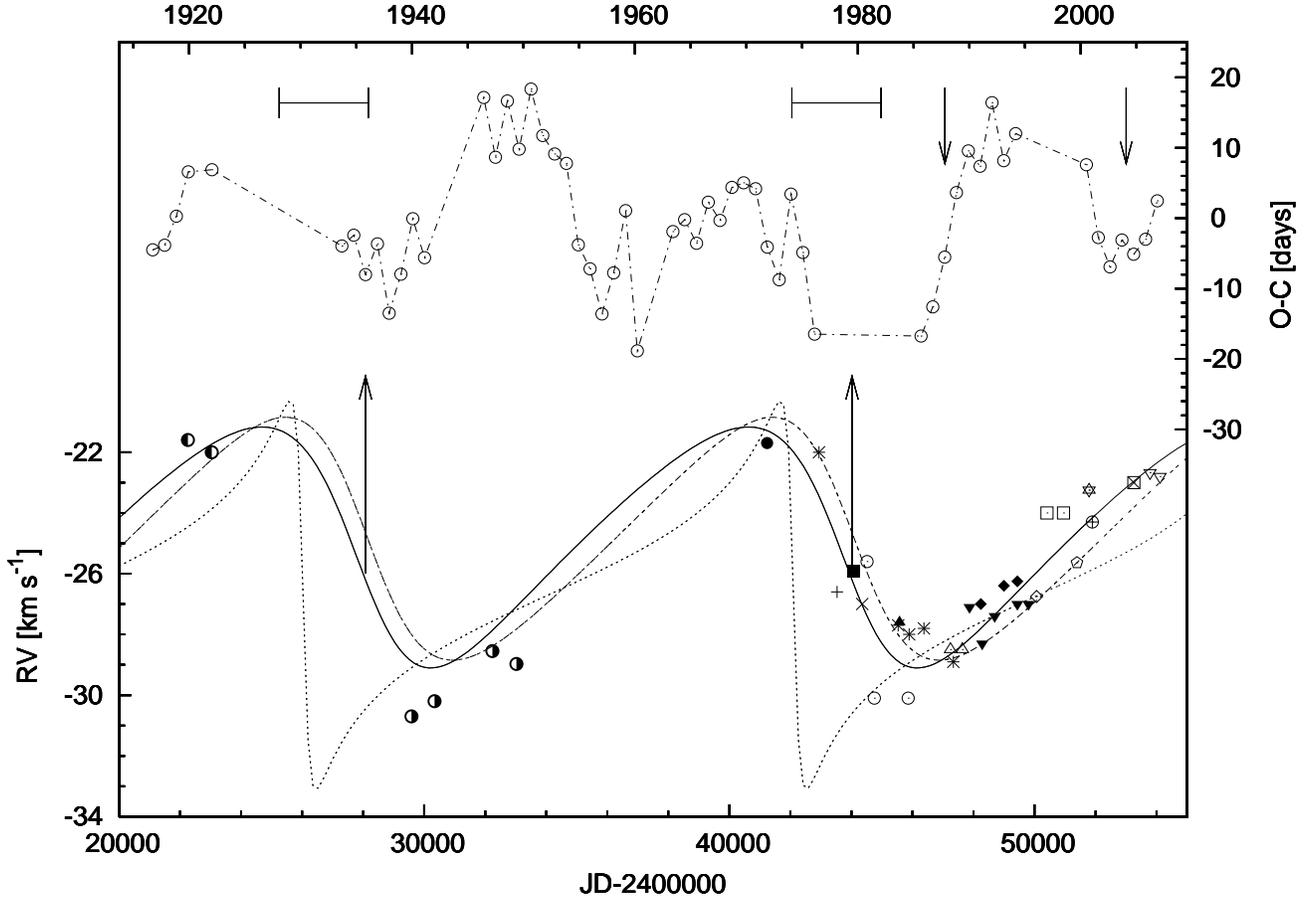}
      \caption{The $O-C$ diagram (top) and  RV variations (bottom) for R Aqr. Observations from different sources are marked like in Table \ref{table:1}. The solid curve represents the fitted orbital elements (Table \ref{table:2}). The dotted curve represents the orbital model proposed by Hollis et al. (\cite{hollis2}). The dashed curve represents solution, with modified $\omega = 87$ which better fits the VLA observations. Bars show times of hypothetic eclipse, short arrows represent moments of jet ejection (Nichols et al. \cite{nichols}), whilst long arrows show periastron passage.
}
         \label{data_jd}
   \end{figure*}

The data from Table 1 of McIntosh \& Rustan (\cite{mcintosh}) were revised and supplemented by additional radial velocity measurements from Merrill (\cite{merrill1}, \cite{merrill2}), 
Zuckerman (\cite{zuckerman}), Martinez et al. (\cite{martinez}), Jewell et al. (\cite{jewell}), Schwarz et al. (\cite{schwarz}), Hollis et al. (\cite{hollis}, \cite{hollis4}), Cotton et al. (\cite{cotton1}, \cite{cotton2}), and Kang et al. (\cite{kang}). 

The blue-violet absorption lines in Mira variables show constant shift of
between $-$5 km s$^{-1}$ and $-$10 km s$^{-1}$ with respect to near-IR velocities (Hinkle et al. \cite{hinkle1}; and references therein).
In our study, a shift equal to $-$6 km s$^{-1}$ was adopted. 
The blue-violet absorption lines do not show any variation with pulsation period
and do not need any additional correction. 
Absorption lines from the red part of the optical spectrum exhibit significant variation but there is no way of correcting for 
this effect in a single measurement.
For this reason, we did not use three data points from Jacobsen \& Wallerstein (\cite{jacobsen}),
which had been previously used for the spectroscopic orbit determinations by Wallerstein (\cite{wallerstein}), 
Hinkle et al. (\cite{hinkle}), and McIntosh \& Rustan (\cite{mcintosh}).

In the near-IR, the radial velocity (RV) variations of R Aqr were dominated by stellar pulsations
($\Delta V \sim 40$ km s$^{-1}$). 
To remove changes due to the pulsations,  
 Hinkle et al. (\cite{hinkle}) complete a linear fit to the infrared velocity curve between phase 0.1 and 0.8, and then adopted as the stellar velocities the residuals remaining after subtracting the velocity-curve fit from each measurement plus the center-of-mass velocity of the Mira of $-29.1\, \rm km\,s^{-1}$. The value of the center-of-mass velocity was estimated from the CO $\Delta \nu=3$ velocity at phase 0.36, following the results of Hinkle, Scharlach \& Hall (\cite{hinkle1}).
Unfortunately, Hinkle et al. (\cite{hinkle}) used an old ephemeris of Campbell (\cite{campbell}). 
Since R~Aqr displays significant variation in its pulsation period (see Fig. \ref{data_jd}), we repeated these corrections with the ephemeris ${\rm{JD(max)}}=2\,442\,404.2 (\pm 0.4)+388.1 (\pm 0.1)\times E$,  
derived for the 1975--1989 period covering Hinkle et al. (\cite{hinkle}) observations.

\begin{table*}
\caption{Radial velocities of R~Aqr used for orbital elements determination.}             
\label{table:1}      
\centering                          
\begin{tabular}{l c c c c c }        
\hline\hline                 
JD & RV $[\rm km\,s^{-1}]$ & Spectral range & Symbol & References \\    
\hline
  2422255 &   -21.6 & Visual & $\LEFTcircle$ & Merrill \cite{merrill1} \\
  2423039 &   -22.0 & Visual & $\LEFTcircle$ & Merrill \cite{merrill1} \\
  2429581 &   -30.7 & Visual &$\RIGHTcircle$ & Merrill \cite{merrill2} \\    
  2430336 &   -30.2 & Visual &$\RIGHTcircle$ & Merrill \cite{merrill2} \\
  2432235.5 & -28.5 & Visual &$\RIGHTcircle$ & Merrill \cite{merrill2} \\
  2433019.3 & -29.0 & Visual &$\RIGHTcircle$ & Merrill \cite{merrill2} \\
  2441237 &   -21.7 & Visual &  $\medbullet$ & Jacobsen \& Wallerstein \cite{jacobsen} \\
  2442939 &   -22.0 & Near-IR & $\divideontimes$ & Hinkle et al. \cite{hinkle} \\
  2443527.5 & -26.6 & $v=1, J=1-0$ & $+$ & Lepine et al. \cite{lepine}, Zuckerman \cite{zuckerman} \\
  2444078 &   -25.9 & $v=1, J=1-0$ & $\blacksquare$& Cohen \& Ghigo \cite{cohen}, Spencer et al. \cite{spencer}, Lane \cite{lane} \\
  2444356 &   -27.0 & $v=1, J=1-0$ & $\times$ & Lane \cite{lane} \\
  2444509 &   -25.6 & Visual &   $\medcirc$ & Wallerstein \cite{wallerstein} \\
  2444748 &   -30.1 & Visual &   $\medcirc$ & Wallerstein \cite{wallerstein} \\
  2445535 &   -27.7 & Near-IR & $\divideontimes$ & Hinkle et al. \cite{hinkle} \\
  2445574.5 & -27.6 & $v=1, J=1-0$ & $\blacktriangle$ & Cho et al. \cite{cho}, Jewell et al. \cite{jewell} \\
  2445862 &   -30.1 & Visual &   $\medcirc$ & Wallerstein \cite{wallerstein} \\
  2445890.6 & -28.0 & Near-IR & $\divideontimes$ & Hinkle et al. \cite{hinkle} \\
  2446378.5 & -27.8 & Near-IR & $\divideontimes$ & Hinkle et al. \cite{hinkle} \\
  2447247 &   -28.5 & $v=1, J=1-0$ & $\triangle$ & Martinez et al. \cite{martinez} \\
  2447338 &   -28.9 & Near-IR & $\divideontimes$ & Hinkle et al. \cite{hinkle} \\
  2447634 &   -28.5 & $v=1, J=1-0$ & $\triangle$ & Martinez et al. \cite{martinez} \\
  2447870 &   -27.1 & $v=1, J=1-0$ & $\blacktriangledown$ & Pardo et al. \cite{pardo} \\
  2448240 &   -27.0 & $v=1, J=2-1$ & $\vardiamondsuit$ & Schwarz et al. \cite{schwarz} \\
  2448283 &   -28.3 & $v=1, J=1-0$ & $\blacktriangledown$ & Pardo et al. \cite{pardo} \\
  2448696 &   -27.4 & $v=1, J=1-0$ & $\blacktriangledown$ & Pardo et al. \cite{pardo} \\
  2448998.4 & -26.4 & $v=1, J=2-1$ & $\vardiamondsuit$ & Schwarz et al. \cite{schwarz} \\
  2449435 &   -27.0 & $v=1, J=1-0$ & $\blacktriangledown$ & Pardo et al. \cite{pardo} \\
  2449439 &   -26.3 & $v=1, J=2-1$ & $\vardiamondsuit$ & Schwarz et al. \cite{schwarz} \\
  2449804 &   -27.0 & $v=1, J=1-0$ & $\blacktriangledown$ & Pardo et al. \cite{pardo} \\
  2450068.8 & -26.8 & $v=1, J=1-0$ & $\diamondsuit$ & Boboltz et al. \cite{boboltz} \\
  2450407 &   -24.0 & $v=1, J=1-0$ & $\Square$ & Hollis et al. \cite{hollis} \\
  2450948 &   -24.0 & $v=1, J=1-0$ & $\Square$ & Hollis et al. \cite{hollis} \\
  2451390.4 & -25.6 & $v=1, J=2-1$ & $\pentagon$ & Kang et al. \cite{kang}, Hollis et al. \cite{hollis} \\
  2451785.8 & -23.2 & $v=1, J=2-1$ & $\davidsstar$ & Kang et al. \cite{kang} \\
  2451892 &   -24.3 & $v=1, J=1-0$ & $\oplus$ & Hollis et al. \cite{hollis4}, Cotton et al. \cite{cotton1}, McIntosh \& Rustan \cite{mcintosh} \\
  2453253 &   -23.0 & $v=1, J=1-0$ & $\XBox$ & Cotton et al. \cite{cotton2} \\
  2453784.3 & -22.7 & $v=1, J=1-0$ & $\triangledown$ & McIntosh \& Rustan \cite{mcintosh} \\
  2454112 &   -22.8 & $v=1, J=1-0$ & $\triangledown$ & McIntosh \& Rustan \cite{mcintosh} \\
\hline                                   
\end{tabular}
\end{table*}

To calculate the orbital solution all data points were averaged in 
387-day bins corresponding to the pulsation period of the Mira component. The velocity changes due to the orbital motion are negligible during such an interval. The data obtained from different methods were binned separately. Since different maser transitions are formed at different
distances from the stellar surface, they were also binned separately.
We excluded single measurements for various maser transitions 
(Allen et al. \cite{allen}; Patel et al. \cite{patel}; Gray et al. \cite{gray}; Imai et al. \cite{imai}) obtained for epochs covered by more extensive data sets.
In Table \ref{table:1}, we list the radial velocity data used to derive the orbital elements of R~Aqr and then converted to local standard of rest ($LSR$) velocities, which are also plotted in Fig. \ref{data_jd}. 

We also decided to exclude from our analysis the SiO maser data obtained during the period JD2\,446\,000--47\,500 (data from Hollis et al. \cite{hollis3}, Heske \cite{heske}, Alcolea et al. \cite{alcolea}, and Hall et al. \cite{hall}) because the maser emission showed at that time significantly different behaviour {\rm e.g.} a few fainter emission, irregular contours than in later epochs, as well as the masers in isolated Miras (see Pardo et al. \cite{pardo}).  
The SiO masers in R Aqr display a ring-like morphology $\sim 31$ mas ($\sim 6.2 [d/200\, \rm pc]$ AU) in diameter (Boboltz et al. \cite{boboltz98}). 
So, these masers lie very close to the stellar surface, namely at a distance of only $\sim 1.5$ radii from the Mira (see Sect. 3) in a region dominated by stellar pulsations and permeated by circumstellar shocks. In addition, the SiO masers can be seriously affected by tidal interactions during the periastron passage when the size of the SiO maser region became comparable to the Roche lobe radius of the Mira.
In fact, according to our orbital solution (see below) the Alcolea et al. (\cite{alcolea}) observations were obtained around periastron.
The deep minimum in both VCs and flux of maser emission that appeared around JD2\,446\,700--47\,300,
coincided with the jet formation (Nichols et al. \cite{nichols}) and a rise in the UV continuum (Meier \& Kafatos \cite{meier}). We suspect that the SiO masers may have been partly disrupted by rising UV radiation during the jet formation, 
and tidal interaction during the periastron passage and then gradually rebuild.

\begin{figure}
   \centering
     \includegraphics[angle=-90,width=9cm]{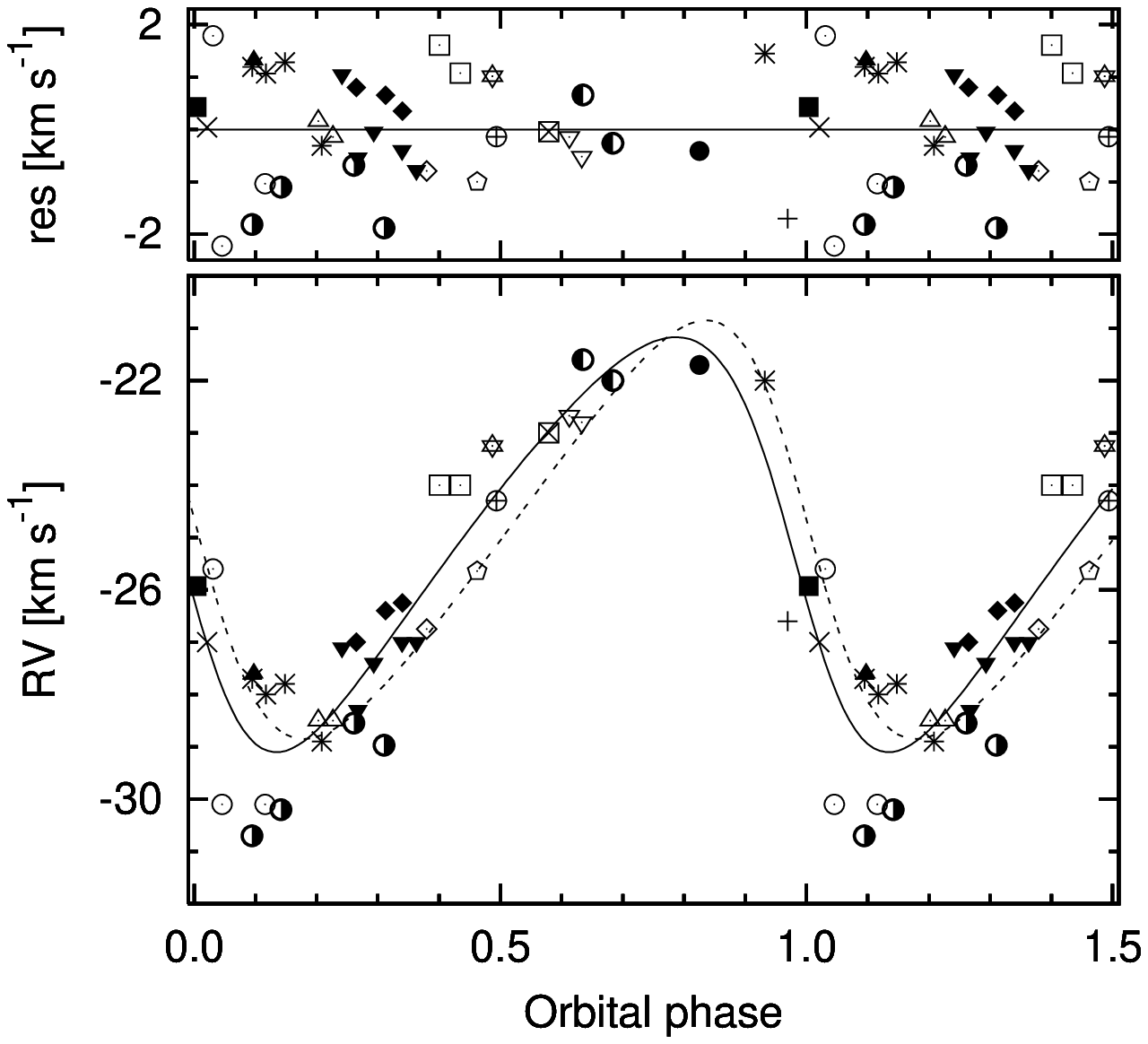}
      \caption{Bottom: The RV data and fitted orbital curve (Table \ref{table:2}) against orbital phase. Symbols are the same as in Fig \ref{data_jd}. The solid curve represent our orbital solution. The dashed curve represents solution, with modified $\omega = 87$ which better fits the VLA observations. Top: Difference between the RV data and our orbital fit (solid line in the lower pannel).
              }
         \label{phased}
   \end{figure}

To identify the spectroscopic orbit, we first estimated a period of RV variations of 43.6 yr using Ortfit (based on the method of Schwarzenberg-Czerny \cite{alex}), and then calculated the orbital parameters: $T_0$, the time of periastron passage, $\gamma$, the baricentral velocity, $K$, the semi-amplitude,
 $\omega$, the longitude of periastron, and $e$, the eccentricity with fixed period. We applied Bertiau's program
(\cite{Bertiau}) based on the Lehmenn-Filh{\'e}s method.
As a result, we obtained an eccentric, $e=0.25\pm0.07$, orbit with reasonable errors (Table \ref{table:2}).
Figure \ref{phased} displays the data and the best-fit velocity curve versus orbital phase.

Figure \ref{data_jd} also compares the RV variations with the $O-C$ data for the Mira pulsations based visual data collected by amateur observers from
AAVSO. 
The $O$--$C$ is a plot showing the difference between the observed  ($O$) times of maximum light and the values calculated 
according to an adopted ephemeris ($C$) as a function of time. A description of the $O$--$C$ method can be found in a review by
Zhou (\cite{zhou}).
To derive a new pulsation ephemeris, we first estimated the
time of each maximum in the light curve
by fitting a third-order Fourier polynomial.
Then a linear ephemeris:
$${\rm{JD(max)}}=2\,416\,070\pm4 + 387.30\pm0.07\times E$$
was fitted to the obtained times of maxima.
The resulting {\it O-C} diagram shows a parabolic shape that
corresponds to an increase in the period rate of $0.013\pm0.002$
day per cycle. After subtracting this parabolic trend,
a sinusoidal variation with a period of 22.5 years and amplitude of about 8 days (Fig. \ref{data_jd}) is visible.

\begin{table}
\caption{Orbital elements for R~Aqr.} 
\label{table:2} 
\centering
\begin{tabular}{l c }
\hline\hline
Element             & Value \\
\hline
$P_{\rm orb}$ (day)    &   15\,943$\pm$471   \\
$\gamma$ (km s$^{-1}$) &   -24.9$\pm$0.2     \\
$K$ (km s$^{-1}$)      &     4.0$\pm$0.4     \\
$e$                    &   0.25$\pm$0.07     \\
$\omega$ (deg)         &      106$\pm$19     \\
$T_{0}$                & 2\,444\,019$\pm$728 \\
$\Sigma (O - C)^2$     &              40     \\
$\sigma$ (km s$^{-1}$) &            1.11     \\
\hline
\end{tabular}
\end{table}

\section{Discussion}

Our orbital solution yields 
the semi-major axis of the Mira of
$$a_{\rm g}\sin{i}=\frac{KP}{2\pi}\sqrt{1-e^2}=5.68^{+0.85}_{-0.83}~{\rm AU},$$
and the mass function of
$$f(\rm M)=\frac{1}{2 \pi G} \cdot P K^3 (1-e^2)^{3/2}=\frac{(M_{\rm h} \sin{i})^3}{(M_{\rm h}+M_{\rm g})^2} = 0.096^{+0.042}_{-0.032} {\rm M}_{\odot},$$ 
where  $M_{\rm g}$, and  $M_{\rm h}$ represent masses of the giant and the hot component, respectively, and the errors are set by the uncertainties in the orbital parameters (see Table \ref{table:2}). The mass function sets a lower limit to the hot companion mass.

\begin{figure}
   \centering
   \includegraphics[angle=-90,width=9cm]{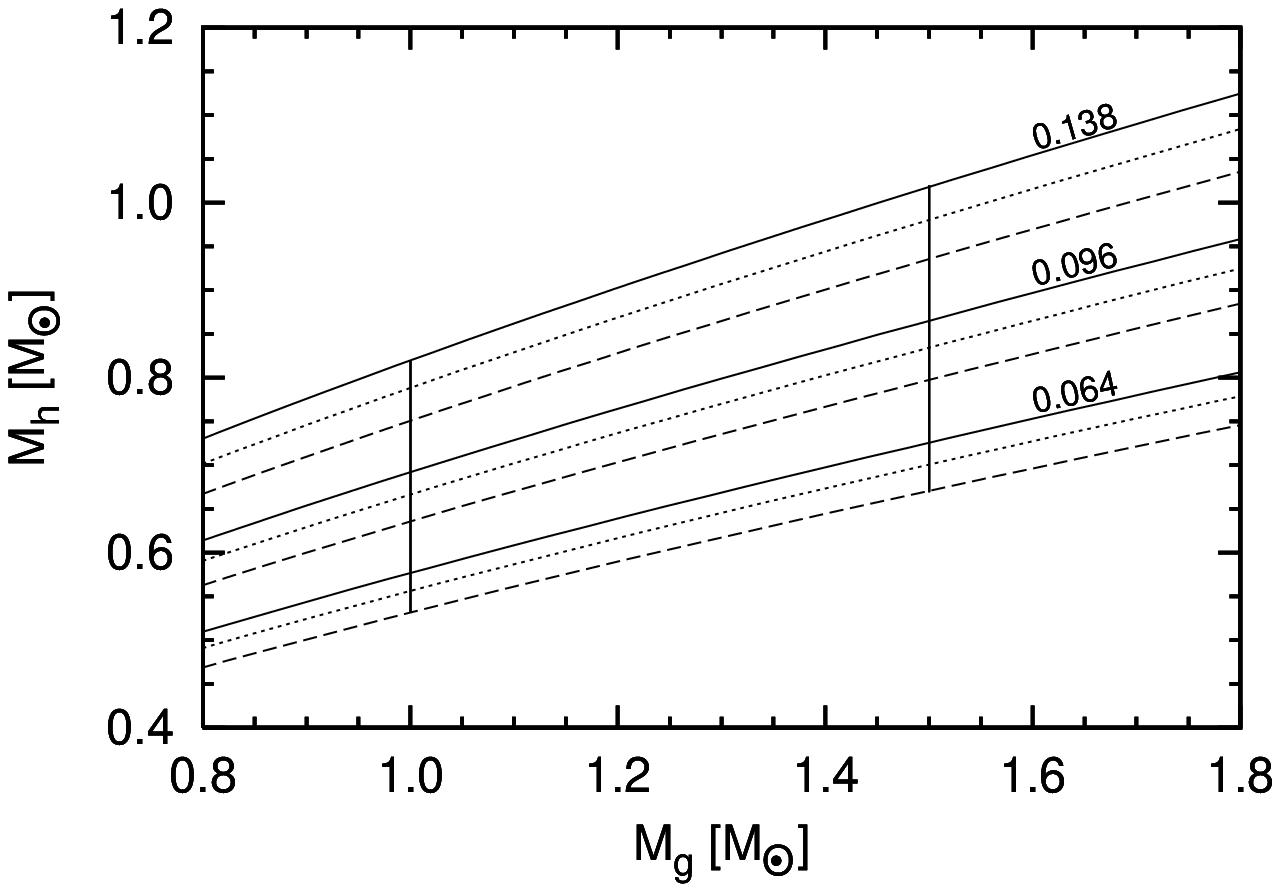}
      \caption{The permitted component masses constrained by the observed mass function  as well as its maximum and minimum values (solid lines labeled with the values of $f(M)$), and assuming $i=70 \degr$. Dotted and dashed lines show the solution for $i=75 \degr$, and $i=90 \degr$, respectively.}
         \label{masses}
   \end{figure}

For further discussion, we assume that the symbiotic binary  R Aqr consist a Mira variable and a white dwarf (WD) companion with an orbit inclined at $i=70\degr$ ($\sim72\degr$ according to Solf \& Urlich \cite{solf}).
Assuming plausible components masses, $M_{\rm g}=$ 1--1.5 M$_{\odot}$, and $M_{\rm h}=$ 0.5--1.4 M$_{\odot}$, for the Mira and the WD, respectively, the total mass of system $M_{\rm tot}=M_{\rm h}+M_{\rm g}$ should be within the range 1.5--3 M$_{\odot}$.
The component masses are also limited by the orbital solution. Therefore, if we consider
$f(M)=$ 0.064--0.138 M$_{\odot}$ and $M_{\rm g}=$ 1--1.5 M$_{\odot}$, we derive $M_{\rm tot}=$ 1.6--2.5 M$_{\odot}$,  $M_{\rm h}=$ 0.57--1.02 M$_{\odot}$, and the mass ratio $q=M_{\rm g}/M_{\rm h}=$ 1.2--2.1 (Fig. \ref{masses}). These values are in good agreement with the commonly adopted model of symbiotic binaries. The semi-major axis of the system, $a$,  is 14.2--16.8 AU, which corresponds to 71--84 mas on the sky for a distance of $d = 200\, \rm pc$ (see below).

Our orbital solution can be tested and refined by using resolved observations of R Aqr (Hege et al. \cite{hege}; Hollis et al. \cite{hollis2}).
In particular, the binary components of R Aqr were resolved by VLA observations (Hollis et al. \cite{hollis2}) obtained on 20 Nov 1996 (JD=2\,450\,407). 
The measured separation between the VLA image in SiO maser $v=1$, $J=1-0$ transition and the continuum emission in a 50 MHz bandwidth at 43.165 GHz associated with an H{\sc ii} region surrounding the hot companion was $\rho=55\pm2\, {\rm mas}$ (11[$d$/200\,pc] AU), and the position angle was $\theta=18\degr\pm2\degr$. 
This can be compared with the projected component separation on the sky plane calculated from our orbital elements.
The orbital elements (Table \ref{table:2}) combined with  the orbital inclination of $i=70\degr$ (Solf \& Urlich \cite{solf}) inferred the projected separation of 
$\rho=0.44a \sim 31$--$37\, [d/200\,{\rm pc}]^{-1}$ mas, and the difference between the apparent binary position angle and the position angle of the line of nodes (the binary orientation on sky) of  $\theta-\Omega = -109 \degr$. 

Our predicted component separation on the sky is consistent with the VLA observations only if the distance is somewhat lower than 200 pc or we underestimate the $\rho$ value due to uncertainties in our orbital elements.

\begin{figure}
   \centering
   \includegraphics[angle=-90,width=9cm]{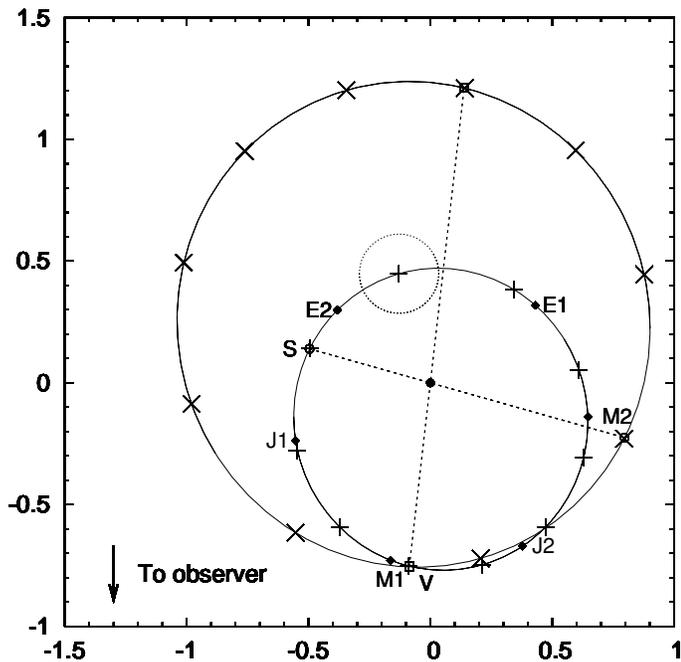}
      \caption{The orbit of the Mira (+) and hot ($\times$)  component in the R~Aqr binary
system in steps of $\Delta\phi = 0.1$. In this representation, the stars  move
anti-clockwise. The dotted circle represents the Mira boundary at $\phi = 0$ (periastron passage). The solid dot marks the mass center. Axes are in units of the semi-major axis $a$. The positions of Mira at the beginning and end of eclipse is marked by E1, and E2, respectively whereas the main and secondary maxima in $O-C$ diagram are denoted by M1 and M2, respectively. J1 and J2 mark the two jet ejection episodes. Open squares connected with dotted line and marked by V show the component position during the VLA observation (Hollis et al. \cite{hollis2}). Open dots connected with dotted line and marked by S show the component position during the speckle observation (Hege et al. \cite{hege}).  }
         \label{orbit}
   \end{figure}

The predicted value of $\rho$ should in fact increase, changing $\omega$ (see also Fig. \ref{orbit}), and/or increasing $e$, and/or decreasing $T_0$. 
For example, $\rho$ should increase to $0.65a$ (46--$55\, [d/200\,{\rm pc}]^{-1}$ mas) just by adopting a lower value of $\omega=  87$, {\rm i.e.} lower only by an amount approximately equal to the standard-deviation uncertainty of $\omega$.

The distance to R~Aqr is also relatively uncertain, and published estimates based on various methods are in the range 180--260 pc (e.g. 180 pc by Solf \& Ulrich \cite{solf}; 181 pc by Lepine et al. \cite{lepine}; 240 pc using the most recent period-luminosity relation from Whitelock et al. \cite{whitelock2008}; and 260 pc by Baade \cite{baade1943}, \cite{baade1944}). The Hipparcos parallax of R~Aqr is  5.07$\pm3.15$ mas (Perryman et al. \cite{perryman}), which corresponds to a distance of $197^{+323}_{-75}$ pc.
Our determination of the parameters of the spectroscopic orbit is thus in a good agreement (within $\leq 1$-$\sigma$ errors in the orbital parameters) with those inferred from the VLA resolved observations.
The value of $\theta=18\degr$, measured with the VLA, implies that $\Omega=127\degr$. Thus, the position of the binary orbit on the sky reproduces well the general picture of the nebula around  R~Aqr, {\rm i.e.} the orbital plane is perpendicular to the jets (see Fig. 7 in Solf \& Urlich \cite{solf}).

The orbital parameters derived for the speckle observations (Hege et al. \cite{hege}) completed on 16 Oct 1983 are inconsistent with our orbit in any case, because the measured separation, $\rho=124 \pm 2$ mas which corresponding to 25[$d$/200\,pc] AU, is too large for any orbital phase. Hege et al. (\cite{hege}) probably did not detect the Mira companion but rather an H$\alpha$ emission region in the SW counterjet.

Our orbital solution predicts the times of spectroscopic conjunctions at
$T_{\rm conj\,I}=2\,450\,859$ (Feb 1998), $T_{\rm conj\,II}=2\,443\,658$
(May 1978) for the inferior ( the Mira is in front of the hot component), and the superior (the Mira behind) conjunction, respectively.
The superior conjuction coincides with the  last dust obscuration/eclipse (see also Fig. \ref{orbit}).
Although the conjunction times could differ by up to 1--2 years due to uncertainties in the orbital parameters, especially $T_{0}$, $e$, and $\omega$, the obscuration lasted about 6 years, and the coincidence between these two events is obvious.

Adopting plausible binary parameters,  i.e. the mass ratio $q\sim1.6$,  $M_{\rm tot}=2$ M$_{\odot}$, and $a=15$ AU (see above), we estimate the minimum component separation  to be $a(1-e) \sim 11 {\rm AU}$,  and the Roche lobe radius for the Mira, to be $R_{\rm L}\sim 4\, {\rm AU}$.
The near-IR interferometry of R~Aqr gave the diameter of the Mira expressed as a diameter of uniform disk, $\Theta_{UD}= 14.06$--20.8 mas in $K$, 17.7 mas in $J$, and 17.7--19.07 mas in $H$, in intermediate  and minimum pulsation phase, respectively (van Belle et al. \cite{vanbelle}; Millan-Gabet et al. \cite{millan}; Ragland  et al. \cite{ragland}). These values correspond to the average radius of the Mira component, $R_{\rm g} \sim 2\, {\rm AU}$.
Even during periastron passage, the Mira variable therefore remains relatively far from filling the Roche lobe.

Many detached interacting binaries, including the symbiotic binaries as well as many X-ray binaries show evidence for much higher mass transfer rate than predicted by models of spherically symmetric winds. This phenomenon can be accounted for by wind focused towards the compact component. Podsiadlowski \& Mohamed (\cite{podsiadlowski}) proposed a model for o~Cet binary system that can explain this focusing. In their model, a slow wind from Mira fills its Roche lobe and then the  matter falls -- via the $L_{1}$ Lagrangian point -- into an accreting stream onto an accretion disk around the companion. This model is even more relevant to R~Aqr where the component separation is significantly smaller than that in o~Cet. In particular, it could explain the increase in extinction towards the Mira  during the last \textquotedblleft eclipse'' in 1974-1981 caused by a neutral material in the accreting stream. According to our orbital solution, the streaming material can obscure the Mira at that time, and in addition, the proximity of periastron  (see Fig. \ref{orbit}) could give rise to enhanced mass transfer.
A similar enhanced, wind-obscuration scenario for R Aqr was discussed by Miko{\l}ajewska \& Kenyon (\cite{mik1}), although in that case the  source of the increased mass loss was proposed to be a helium flash above a massive core ($M_{\rm core}\sim1.36 M_{\odot}$) of the Mira.

The orbital configuration of R~Aqr 
could also explain the observations of the OH and H$_2$O masers that originate in circumstellar regions more distant from the Mira variable ({\rm e.g.} H$_2$O masers zone is 10--20 AU)  than the SiO maser emission region.
The OH and H$_2$O maser lines were not detected in 1979 and 1984 (Cohen \& Ghigo \cite{cohen}; Norris et al. \cite{norris}) when the hot companion was in front of Mira. 
Ivison et al. (\cite{ivison}) reported detection of weak OH and H$_2$O maser emission in 1993, when the Mira was in front of the hot companion and the masers were shielded from its radiation. 
They claimed, however, that the OH maser detection was only tentative, and that the velocity measurement was very unusual (an expansion velocity of 10 km s$^{-1}$ is rather large for this system), and only the red peak should have been detected, and additional  spectra obtained using the VLA during 1994 June did not show the OH maser line (Ivison et al. \cite{ivison98}).
The  H$_{2}$O line appeared more promising, although weak for a Mira variable with relatively strong SiO masers.
It was also clearly detected during 3 epochs in 1993--1995 (Ivison et al. \cite{ivison98}). 
Ivison et al. (\cite{ivison98}) proposed three possible mechanisms that could inhibit the OH and H$_2$O maser formation in their natural environment, involving either the orbital motion of the companion or its UV radiation and fast wind, or line obscuration at low frequencies by optically thick, ionized gas.

The apparent correlation between the orbital position and the $O-C$ for the Mira pulsations is extremely interesting. In particular, minima in the $O-C$ diagram occur around the periastron (see Fig \ref{data_jd}). This may be due to some distortion of the Mira caused by tidal interaction during the periastron passage, for example if the Mira became
elongated towards the hot component, the pulsation may need more time for propagation in this direction. 
According to our orbital model, the main maximum of $O-C$ appears close to the inferior conjunction, when the Mira eclipses the hot companion, whereas the secondary, lower maximum appears in the middle of the two conjunctions, when both components are well separated and we can see the elongated side.

Finally,  the orbital solution proposed is inconsistent with the 17 yr orbital period for R~Aqr proposed by Nichols et al. (\cite{nichols}). They noticed that
radio observations obtained in 1987 and 2004 exhibited non-thermal spectra, which are usually interpreted to be a signature of a newly emitted  jet (in stellar sources older jets are generally thermal), suggested 
that the jet formation is coupled with periastron. In our opinion, such a short period appears unrealistic not only because it is not reflected by the RV variation but also because in that case there would not be sufficient room for a Mira with a dust envelope. The 17-yr period implies a semi-major axis of $\sim8$ AUa whereas the minimum component separation for D-type symbiotic binaries is about 10--15 AU,
which for typical masses (see above) indicates orbital periods longer than 20 yrs (Miko{\l}ajewska \cite{mik2}). It appears that the jet ejection in R~Aqr connot be connected with the orbital motion.

\section{Conclusions}

Based on published radial velocities of its Mira component, we have derived new orbital parameters for the symbiotic binary R~Aqr. We have found in particular, the orbital period of 43.6 yr, and showed that the mass function is consistent with the presence of a typical, 1--1.5 $\rm M_{\odot}$, Mira variable accompanied by a 0.6--1 $\rm M_{\odot}$ white dwarf. We also showed that our spectroscopic orbit is consistent  with the VLA astrometry (Hollis et al. \cite{hollis2}) and inconsistent with the
speckle interferometery (Hege et al. \cite{hege}).
Our orbital model allow us to interpret the \textquotedblleft eclipses'' of Mira as obscuration by the accretion stream.
We also showed that the jet ejection (Nichols et al. \cite{nichols}) is not connected with the orbital position.
Finally, we note that September 2012 should be a perfect time to complete resolved imaging of R~Aqr, since the Mira component will then be at minimum brightness and the separation
between the components on sky will be close to its maximum possible value.

\begin{acknowledgements}
      This study made use of the American Association of Variable Star Observers (AAVSO) International Database contributed by observers worldwide. We are grateful to the anonymous referee for valuable comments.
      We  also thank Radek Smolec for comments to the first version of this paper
      and Wojtek Pych for providing the Ortfit software.
      This work was partly supported by the Polish Research grant N203\,395534.
\end{acknowledgements}

\end{document}